\documentclass{nature}
\usepackage{epsfig}
\usepackage{amsmath}
\usepackage{amssymb,color}
\usepackage{leftidx}
\usepackage{color}
\usepackage{mathrsfs}

\begin{document}

\title{Topological Superfluidity of Spin-Orbit Coupled
Bilayer Fermi Gases}
\author{Liang-Liang Wang, Wu-Ming Liu$^{\star}$}
\maketitle
\begin{affiliations}

\item
Beijing National Laboratory for Condensed Matter Physics, Institute of Physics, Chinese Academy of Sciences, Beijing 100190, China

$^\ast$e-mail: wliu@iphy.ac.cn

\end{affiliations}

\begin{abstract}
Topological superfluid, new quantum matter that possesses gapless exotic excitations known as Majorana fermions, has attracted extensive attention recently. 
These excitations, which can encode topological qubits, could be crucial ingredients for fault-tolerant quantum computation. 
However, creating and manipulating multiple Majorana fermions remain an ongoing challenge. 
Loading a topologically protected system in multi-layer structures would be a natural and simple way to achieve this goal. 
Here we investigate the system of bilayer Fermi gases with spin-orbit coupling and show that the topological condition is significantly influenced by the inter-layer tunneling, yielding two novel topological phases, which support more Majorana Fermions. 
We demonstrate the existence of such novel topological phases and associated multiple Majorana fermions using bilayer Fermi gases trapped inside a harmonic potential. 
This research pave a new way for generating multiple Majorana fermions and would be a significant step towards topological quantum computation.
\end{abstract}

Topological state of matter \cite{TopologicalSuperfluid-2010Kane,TopologicalSuperfluid-2011XiaoLiang}, as an active new frontier in physics, has been attracting considerable interest both experimentally and theoretically. 
The so-called Majorana fermions (MFs) \cite{MajoranaFermion-2011Zoller,MajoranaFermion-2012Alicea,MajoranaFermion-2012Brouwer,MajoranaFermion-2015Franz}, accompanied with topological states, have a good prospect for applications in topological quantum computation, quantum memory and quantum random-number generation \cite{QuantumComputer-2001Kitaev,QuantumComputer-2008Nayak,QuantumComputer-2013DongLing} since they are intrinsically immune to decoherence caused by local perturbations. 
Due to such potential applications, the realization of topological states in a well-controlled environment is highly desirable. 
Recently, several theoretically predictions and experimentally observations of topological states have been reported in a variety of systems, for example, InAs wires and banded carbon nanotubes engineered by spin-orbit coupling of electrons \cite{InAsAndNanoTube-2010Alicea,InAsAndNanoTubeAndGapClose-2010Lutchyn,InAsAndNanoTubeAndGapClose-2010Oreg,InAsAndNanoTube-2011Stanescu,InAsAndNanoTube-2011Stoudenmire}, the surface of topological insulators \cite{TopologicalInsulator-2008Fu,TopologicalInsulator-2008Liu,TopologicalInsulator-2010Alicea,TopologicalInsulator-2010Chen,TopologicalInsulator-2010Sau} , ultracold atomic gases with strong non-Abelian synthetic gauge field \cite{UltracoldAtomicGases-2007Tewari,UltracoldAtomicGases-2011Vyasanakere,UltracoldAtomicGases-2012Gong,UltracoldAtomicGases-2012Liu,UltracoldAtomicGases-2012Zhai,UltracoldAtomicGases-2013Fu,UltracoldAtomicGases-2013Qu,UltracoldAtomicGases-2013Zhang1,UltracoldAtomicGases-2013Zhang2,UltracoldAtomicGases-2014Buhler,UltracoldAtomicGases-2014Hamner} and so on. 
Spin-orbit (SO) coupling plays a key role in all systems. 
In contrast to the other physical contexts, ultracold atomic gases with more experimental controls over its parameters, such as high controllability in interatomic interaction, geometry, purity, etc, offer an exceptional clean platform for exploring the topological superfluidity. 

Recent experimental progresses in the synthetic SO coupling of ultracold Bose gases \cite{BoseExper-2009Lin1,BoseExper-2011Lin2} and Fermi gases \cite{FermiExper-2011Sau,FermiExper-2012Zhangjing,FermiExper-2012Zwierlein} have stimulated a lot of interest in exploring exotic topological quantum properties of ultracold atoms. 
In particular, by deforming the Fermi surface and opening a topological band gap, SO coupled fermi gases become topological and support Majorana zero-energy excitations in the presence of a large perpendicular Zeeman field. 
However, based on these experiments, the existence of a topological band  gap can only lead to a single topological phase and a pair of Majorana zero modes associated with topological quantum phase transition (TQPT). 
There is a strong requirement in the realm of quantum information and quantum memory for proposing a advanced system, which results in more types of topological phases, to create and manipulate multiple MFs. 
To achieve this, we give a realistic scenario based upon engineering a topologically protected system in a bilayer or multi-layer structures. 
As we know, some intriguing quantum effects, generated by layers's extra degrees of freedom, have been widely studied in many multi-layer condensed matter systems including fractional quantum Hall states \cite{BilayerSystemofCondensedMatter-2010Peterson} and topological valley transport in bilayer graphene \cite{BilayerSystemofCondensedMatter-2015Sui}. 
Interesting quantum effects such as atomic Josephson effects and macroscopic quantum self-trapping \cite{BilayerBEC-1997Milburn,BilayerBEC-1997Smerzi,BilayerBEC-2000Giovanazzi,BilayerBEC-2005Albiez,BilayerBEC-2007Levy,BilayerBEC-2011LeBlanc} have also been studied in ultracold Bose gases. 
However, for ultracold Fermi gases, most of the investigations have been developed mainly in the field of single-layer topological system\cite{SpinorBose-2010Wang,SpinorBose-2011Ho,SpinorBose-2012Wen,SpinorBose-2013Liu}, very little is known regarding the bilayer Fermi gases with SO coupling. 
It would be interesting to examine the possibility for realizing and manipulating multiple MFs by loading SO coupled Fermi gases in a bilayer structure, providing a peculiar insight into the topological states. 

In this article, we concentrate on the potential topological properties of bilayer Fermi gases with SO coupling and predict the existence of multiple Majorana zero states at the interfaces between distinct phases. 
The bilayer geometry can be readily realized by adding a two-dimensional double-well optical lattice, and the atoms in each layer are affected by the same Raman SO coupling, while coupled with each other via the inter-layer tunneling. 
Notably different from previous mean field studies, by tunning the inter-layer tunneling, traditional topological phase condition induced by the collective effect of SO coupling and Zeeman field is shifted and causes two critical transition points [$h_{c,1}$,$h_{c,2}$]. 
As a consequence, the topological phase region becomes separated into two parts: topo-\uppercase\expandafter{\romannumeral1} and topo-\uppercase\expandafter{\romannumeral2} phases. 
Here we provide a physical picture to classify the two different topological phases and discover that in suitable parameter regions, the system may undergo a unique TQPT from a topo-\uppercase\expandafter{\romannumeral1} state to a topo-\uppercase\expandafter{\romannumeral2} state, which has not been previously identified. 
The region for topological superfluid depends on not only magnitude of the Zeeman field, the chemical potential, the SO coupling strength, but also on the inter-layer tunneling, thus provides more knobs in experiments. 
Furthermore, we give the self-consistent Bogoliubov-de Gennes (BdG) results of the system trapped in a harmonic potential and discuss the resulting multiple Majorana zero-energy modes in great detail. 
The number of exotic Majorana zero modes depends on the number of interfaces between distinct phases that form in different areas of the trap, which are consistent with analytical local-density approximation (LDA) expressions. 
Such SO coupled bilayer Fermi gases offer an opportunity to create and manipulate multi-MFs and a advanced multi-layer experimental setup may be therefore improved to measure more MFs.
\section*{Results}
\subsection{Experimental setup of spin-orbit coupled bilayer Fermi gases.}
SO coupling for ultracold Fermi atoms has been successfully demonstrated in ultracold Fermi $\leftidx{^{40}}{\mathrm{K}}$ and $\leftidx{^{6}}{\mathrm{Li}}$ gases \cite{FermiExper-2012Zwierlein,FermiExper-2012Zhangjing} at about the same time, in which the Raman dressing scheme is based on coupling two magnetic sub-levels of the ground state manifold with two counter propagating Raman lasers \cite{BoseExper-2009Lin1,FermiExper-2011Sau}. 
The system considered in this work is depicted in Fig. 1(a), where the Raman dressed $\leftidx{^{40}}{\mathrm{K}}$ Fermi gases are loaded in a bilayer geometry, which can be readily realized by adding a two-dimension double-well optical lattice \cite{QuantumDoubleWell-2006Strabley,QuantumDoubleWell-2015Xu}. 
Note that the optical lattice is spin-independent and can induce tunneling without spin-flip. 
Same two atomic internal spin states
$|9/2,9/2\rangle$ and $|9/2,7/2\rangle$ of the ground state in $j$-th layer are selected to be labeled as spin-up ($|j,\uparrow\rangle$) and down ($|j,\downarrow\rangle$) states, where $j=1,2$ refers to an individual layer. 
The atom move along the $\hat{x}$ axis within a layer and two layers are separated by a distance $d$. 
The tunneling amplitude $J_{\sigma}=\int dz\psi_{1,\sigma}^{\ast}(z)[V(z)\pm\delta]\psi_{2,\sigma}(z)$ with $V(z)=c(z^{2}-d^{2})^{2}$ can be regulated by changing the intensity or relative phase of laser standing waves that engineer the double-well lattice \cite{QuantumDoubleWell-2006Strabley}. 
The positive and negative signs label the detuning from Raman resonance to spin-up and spin-down respectively. 
Without loss of generality, we assume them to be real and $J_{\uparrow}=J_{\downarrow}=J$. 
The four-level topology has been schematically shown in Fig. 1(b), where the blue solid lines represent the inter-layer tunneling and the green dashed circles denote the Raman-assisted intra-layer interaction with momentum transfer $k_{0}$. 
The inter-layer spin-flip tunnelings are negligible under current experimental conditions, therefore they are neglected in Fig. 1(b). 
Four atomic states couple with each other in a cyclic manner with no momentum transferred during a closed loop transition. 

The Hamiltonian for bilayer Fermi gases in the presence of SO coupling is given by $H=\int dx [\mathscr{H}_{S}+\mathscr{H}_{\mathrm{int}}]$, with the single-particle component $\mathscr{H}_{S}$ and the interacting component $\mathscr{H}_{\mathrm{int}}= \sum_{j=1,2}g_{j}\Psi^{\dagger}_{j\uparrow}(x)\Psi^{\dagger}_{j\downarrow}(x)\Psi_{j\downarrow}(x)\Psi_{j\uparrow}(x)$ describing the $s$-wave contact interaction between the two spin states in $j$-th layer. 
The single-particle Hamiltonian is written as
\begin{equation}
\begin{split}
\mathscr{H}_{S}=\Psi^{\dagger}(x)
\begin{pmatrix}
& \xi_{k}+V(x)+\delta & \Omega e^{ik_{0}x} & J & 0 \\
& \Omega e^{-ik_{0}x} & \xi_{k}+V(x)-\delta & 0 & J \\
& J & 0 & \xi_{k}+V(x)+\delta & \Omega e^{ik_{0}x} \\
& 0 & J & \Omega e^{-ik_{0}x} & \xi_{k}+V(x)-\delta\\
\end{pmatrix}
\Psi(x)
\end{split}
,
\end{equation}
with $\Psi(x)=[\Psi_{1,\uparrow}(x),\Psi_{1,\downarrow}(x),\Psi_{2,\uparrow}(x),\Psi_{2,\downarrow}(x)]^{T}$ being the atomic annihilation operators, $\xi_{k}=\epsilon_{k}-\mu$, where $\epsilon_{k}=k_{x}^{2}/2m$ denotes kinetic energy and $\mu$ is the chemical potential, $m$ is mass of an atom, $\delta$ means the detuning of the Raman process, and $V(x)$ is the trapping potential (for convenience, we set $\hbar=k_{B}=1$). 
The constants $\Omega$ and $k_{0}$ represent the coupling strength and photon recoil momentum of the two-photon Raman coupling, respectively. 
After applying a local gauge transformation
\begin{equation}
\begin{split} & \Psi_{j,\uparrow}(x)=\frac{1}{\sqrt{2}}[e^{ik_{0}x/2}\Phi_{j,\uparrow}(x)-ie^{ik_{0}x/2}\Phi_{j,\downarrow}(x)], \\ & \Psi_{j,\downarrow}(x)=\frac{1}{\sqrt{2}}[e^{-ik_{0}x/2}\Phi_{j,\uparrow}(x)+ie^{-ik_{0}x/2}\Phi_{j,\downarrow}(x)],  \\
\end{split}
\end{equation}
the single-particle Hamiltonian becomes
\begin{equation}
\mathscr{H}_{S}= \sum_{j}\Phi^{\dagger}_{j}(x)H_{0}\Phi_{j}(x)+J\sum_{\sigma}(\Phi^{\dagger}_{1,\sigma}(x)\Phi_{2,\sigma}(x)+\mathrm{H.c.}),
\end{equation}
where $\Phi_{j}(x)=[\Phi_{j,\uparrow}(x),\Phi_{j,\downarrow}(x)]^{T}$ is a single layer field operator, and the single layer term is
\begin{equation}
H_{0}=\epsilon_{k}-\mu+V(x)+(\alpha k_{x}+h_{y})\sigma_{y}-h_{z}\sigma_{z},
\end{equation}
where we define the chemical potential $\mu\rightarrow \mu-E_{r}/4$ and $\sigma_{y}$, $\sigma_{z}$ are the Pauli matrices acting on the spins. 
The effective spin-orbit coupling constant $\alpha\equiv E_{r}/k_{r}$ and effective Zeeman field ($h_{y}\equiv \delta$, $h_{z}\equiv\Omega$) are introduced. 
For convenience, the recoil momentum $k_{r}\equiv k_{0}$ and recoil energy $E_{r}\equiv k_{0}^{2}/2m$ are taken as natural momentum and energy units.
Notice that the interaction Hamiltonian $\mathscr{H}_{\mathrm{int}}$ is invariant under this gauge transformation \cite{InteractionInvariant-2015Liu}. 

We first consider homogeneous Fermi gases to give the simplest qualitative picture of the bilayer system. 
At this point, the physics of the single-particle Hamiltonian is simple in momentum space using a Fourier decomposition. 
The four-band structure of the bilayer system is illustrated in Fig. 1(c,d). 
Obviously, the in-plane Zeeman field $h_{y}$ leads to a Fermi surface without inversion symmetry and gives a so-called Fulde-Ferrell-Larkin-Ovchinnikov (FFLO) state with a finite total momentum pairing \cite{UltracoldAtomicGases-2013Qu,UltracoldAtomicGases-2013Zhang2}, but can not open a band gap at the Dirac point, as shown in Fig. 1(d). 
For the realization of topological phase, a perpendicular Zeeman field is needed to break the time-reversal (TR) symmetry and open a topological band gap. 
In Fig. 1(c), we report that the topological band gaps occur at the Dirac point $\mathbf{k}=0$ induced by increasing $h_{z}$, which are similar to the single-layer mechanism. 
The superfluid pairing is between atoms with opposite momentum $-\mathbf{k}$ and $\mathbf{k}$ without in-plane Zeeman field $h_{y}$. 
A critical observation is that when the Zeeman field is larger than the tunneling effect $h_{z}>J$, the increasing Zeeman field removes the level crossing and open a gap between the two middle branches, as shown by the red curves in the Fig. 1(c,d). 
We will give a detailed discussion in the mean-field formalism. 

\subsection{Classification of topological superfluid phases.}
We consider the most simplified form of interactions with superfluid pairing formed between atoms with opposite momentum ($-\mathbf{k}$, $\mathbf{k}$). 
Within the mean-field approximation, the interaction term can be rewritten as
\begin{equation}
\mathscr{H}_{\mathrm{int}}\approx \sum_{j=1,2}\sum_{k}\Delta_{j}\Phi^{\dagger}_{j,k,\uparrow}\Phi^{\dagger}_{j,-k,\downarrow}+\Delta_{j}^{\ast}\Phi_{j,-k,\downarrow}\Phi_{j,k,\uparrow}-\frac{|\Delta_{j}|^{2}}{g_{j}},
\end{equation}
with the order parameters $\Delta_{j}=g_{j}\sum_{k}\langle\Phi_{j,-k,\downarrow}\Phi_{j,k,\uparrow}\rangle$ . 
One can easily write the mean-field BdG Hamiltonian,
\begin{equation}
H_{\mathrm{BdG}}(k)=
\begin{pmatrix}
& H_{S}^{0}(k) & J & 0 & \sigma_{z}\Delta_{1}\sigma_{x} \\
& J & H_{S}^{0}(k) & \sigma_{z}\Delta_{2}\sigma_{x} & 0 \\
& 0 & -\sigma_{z}\Delta_{2}\sigma_{x} & -\sigma_{z}H_{S}^{0}(-k)\sigma_{z} & -J \\
& -\sigma_{z}\Delta_{1}\sigma_{x} & 0 & -J & -\sigma_{z}H_{S}^{0}(-k)\sigma_{z} \\
\end{pmatrix}
,
\end{equation}
where the Nambu spinor basis is chosen as $[c_{k,1,\uparrow}, c_{k,1,\downarrow}, c_{k,2,\uparrow}, c_{k,2,\downarrow}, c^{\dagger}_{k,2,\uparrow}, c^{\dagger}_{k,2,\downarrow}, c^{\dagger}_{k,1,\uparrow}, c^{\dagger}_{k,1,\downarrow}]^{T}$.
The elementary excitations can be found by solving the
BdG equation $H_{\mathrm{BdG}}(k)W_{\eta}^{\pm}(k)=E_{\eta}^{\pm}(k)W_{\eta}^{\pm}(k)$, where $E_{\eta}^{\pm}(k)$ ($\eta=1,2,3,4$) are the eigenvalues of the above $8\times 8$ BdG Hamiltonian and corresponding wave functions are assumed as $W_{\eta}^{\pm}(k)=(u_{1,\uparrow,\eta}^{\pm}, u_{1,\downarrow,\eta}^{\pm}, u_{2,\uparrow,\eta}^{\pm}, u_{2,\downarrow,\eta}^{\pm},v_{2,\uparrow,\eta}^{\pm}, v_{2,\downarrow,\eta}^{\pm}, v_{1,\uparrow,\eta}^{\pm}, v_{1,\downarrow,\eta}^{\pm})^{T}$ , where the label $\pm$ represent the particle as well as hole branches. 
Without loss of generality, we set $\Delta=\Delta_{1}=\Delta_{2}$ throughout the work. 
The spectrum of $H_{\mathrm{BdG}}(k)$ consists of eight bands given by
\begin{equation}
E_{\eta=1,2}^{\pm}(k)=\pm\sqrt{b_{1}(k)\pm 2\sqrt{d_{1}(k)}}, E_{\eta=3,4}^{\pm}(k)=\pm\sqrt{b_{2}(k)\pm 2\sqrt{d_{2}(k)}},
\end{equation}
where the $b_{1}(k)=(k^{2}/2m-\mu-J)^{2}+\Delta^{2}+h_{z}^{2}+\alpha^{2}k^{2}$ and $d_{1}(k)=(k^{2}/2m-\mu-J)^{2}(\alpha^{2}k^{2}+h_{z}^{2})+\Delta^{2}h_{z}^{2}$ with $b_{2}(k)=(k^{2}/2m-\mu+J)^{2}+\Delta^{2}+h_{z}^{2}+\alpha^{2}k^{2}$ and $d_{2}(k)=(k^{2}/2m-\mu+J)^{2}(\alpha^{2}k^{2}+h_{z}^{2})+\Delta^{2}h_{z}^{2}$.
By solving $E_{\eta}^{+}(0)=0$, we obtain all the gap close conditions in analytic forms:
\begin{equation}
h_{z}^{4}-C_{1}h_{z}^2+C_{2}=0,
\end{equation}
with $C_{1}=2(\Delta^2+J^2+\mu^2)$, $C_{2}=\Delta^4+J^4+\mu^4+2\Delta^2J^2+2\Delta^2\mu^2-2J^2\mu^2$. The solutions for the Zeeman field are
\begin{equation}
h_{z}=\sqrt{\Delta^2+(J\pm \mu)^2},
\end{equation}
here we just consider the parallel case $h_{z}\geqslant0$. If there is no tunneling between two layers, the critical Zeeman field reduces to the well-known $h_{c}=\sqrt{\mu^2+\Delta^2}$, which has been discussed in detail in earlier works \cite{InAsAndNanoTubeAndGapClose-2010Oreg,InAsAndNanoTubeAndGapClose-2010Lutchyn}.
The system will be in a conventional superfluid at $h_{z}<h_{c}$ and in a topological superfluid at $h_{z}>h_{c}$.
Unlike the conventional phase transition in a single layer system, the tunneling shifts the chemical potential and gives two critical Zeeman fields: $h_{c,1}=\sqrt{\Delta^{2}+(J-\mu)^{2}}$ and $h_{c,2}=\sqrt{\Delta^{2}+(J+\mu)^{2}}$ (or exchange to assure $h_{c,1}<h_{c,2}$). The system undergoes a TQPT when the Zeeman field cross these two critical values, where the single-particle excitation gap vanishes, representing a topological phase transition. To see the phase transitions more clearly, we give the spin population at $k=0$, as shown in Fig. 2(b). 
Here we define the spin vector expectation value along the $z$-axis as
\begin{equation}
\begin{split}
\langle S_{z}(k)\rangle &=\sum_{j}\langle\Phi^{\dag}_{j,k,\uparrow}\Phi_{j,k,\uparrow}-\Phi^{\dag}_{j,k,\downarrow}\Phi_{j,k,\downarrow}\rangle\\
&=\sum_{j,\eta}\langle|v_{j,\uparrow,\eta}^{+}|^{2}-|v_{j,\downarrow,\eta}^{+}|^{2}\rangle.\\
\end{split}
\end{equation}
We find that $\langle S_{z}(0)\rangle$ changes discontinuously when the Zeeman field crosses the critical values $[h_{c,1}, h_{c,2}]$, which implies the change of the topology of the spin texture. 
$\langle S_{z}(0)\rangle=0$ if $h_{z}<h_{c,1}$, which corresponds to the trivial state, while $\langle S_{z}(0)\rangle=1$ for $h_{c,1}<h_{z}<h_{c,2}$ and $\langle S_{z}(0)\rangle=2$ for $h_{z}>h_{c,2}$ denote diverse topological states, as show in Fig. 2(b). 
Therefore we give a classification of topological superfluid phases according to the strength of perpendicular Zeeman field $h_{z}$:
\begin{equation}\label{equ:GapCloseCondition}
\begin{split}
\mathbf{trivial\; region}: & 0<h_{z}<h_{c,1},\ \Delta\neq0,\\
\mathbf{topo-\uppercase\expandafter{\romannumeral1}\; region}: & h_{c,1}<h_{z}<h_{c,2},\ \alpha \Delta\neq0,\ E_{g}>0,\\
\mathbf{topo-\uppercase\expandafter{\romannumeral2}\; region}: & h_{z}>h_{c,2},\ \alpha \Delta\neq0,\ E_{g}>0,\\
\end{split}
\end{equation}
where the last condition $E_{g}=\min\{E^{+}_{j,\sigma}(k)\}>0$ in the topological regimes ensures the bulk quasi-particle excitations are gapped to protect the zero-energy MFs. The region for topological superfluid depends not only on the chemical potential, pairing strength, but also on the tunneling strength. It provides more control knobs for tuning the topological phase transition. Especially when the tunneling can be comparable to the chemical potential, the perpendicular Zeeman field threshold can be dramatically lowered. These new features may open a possibility for the experimental realization of topological superfluid under small Zeeman field.

\subsection{Phase diagram of spin-orbit coupled bilayer Fermi gases.}
To better understand the transition from one state to another defined by equation (\ref{equ:GapCloseCondition}), it is necessary to observe the close and reopen of the excitation gap $E_{g}$, which change the topology of Fermi surface.
The phase diagram in the plane of the order parameter and the perpendicular Zeeman field is presented in Fig. 2(a). The graph is colored according to the energy gap $E_{g}$, and white dotted curves mark contours of $E_{g}=0$.
There are three different topological regions determined by the behavior of the energy gap.
The system is topological trivial at first, as increasing the Zeeman field with a fixed $\Delta$, the band gap may first close and then reopen, signifying the transition from non-topological to $\mathrm{topo-\uppercase\expandafter{\romannumeral1}}$ superfluid ($h_{z}\in[h_{c,1},h_{c,2}]$), and finally undergoes a topological phase transition from $\mathrm{topo-\uppercase\expandafter{\romannumeral1}}$ to $\mathrm{topo-\uppercase\expandafter{\romannumeral2}}$ phase at the second gapless point $h_{c,2}$.
In Fig. 2(b-f), we plot the band gap and energy spectrums to illustrate the TQPT when $\Delta=0.5E_{r}$.
It should be emphasized that the range of region: $\mathrm{topo-\uppercase\expandafter{\romannumeral1}}$ is highly dependent on the value of order parameter $\Delta$, the increasing of the intensity of $\Delta$ rapidly shrinks this region.

As have discussed the evolution of the single-particle Hamiltonian as a function of $h_{z}$, here we can further more rigorously demonstrate the topological phase transition with the change of the chemical potential $\mu$.
We can sketch the phase diagram in the Fig. 3(a), which shows perfect inversion symmetry ($\mu \rightarrow -\mu$) in the $\mu-\Delta$ plane.
For a small order parameter, the important feature is that there have four topological transition points along the change of the chemical potential.
Similar to the case in Fig. 2, by tuning through the transition point, the quasi-particle excitation gap closes and open again, thus we can discriminate five regions in accordance with the topological condition (\ref{equ:GapCloseCondition}) with central area being $\mathrm{topo-\uppercase\expandafter{\romannumeral2}}$.
We can write down the scaling forms of the critical chemical potentials
\begin{equation}
\mu=\pm(\pm\sqrt{h_{z}^{2}-\Delta^{2}}-J),
\end{equation}
which can be easily got from the gap closing condition (9).
It should be pointed out that the range of central region $\mathrm{topo-\uppercase\expandafter{\romannumeral2}}$ gets shrink by increasing $\Delta$ and turn into trivial phase above a threshold: $\Delta^{2}=h_{z}^{2}-J^{2}$.
By further increasing the intensity of the order parameter beyond $\Delta=h_{z}$, the inequality $h_{z}-\sqrt{\Delta^{2}+(J\pm \mu)^{2}}<0$ holds at any $\mu$, thus there is only trivial phase region in this situation.
The evolvement of band gap $E_{g}$ is plotted as the change of chemical potential in Fig. 3(b).
There are five phases and separated by the critical values of the chemical potential $\mu_{c,i}$ with $i=1,2,3,4$.
It can be better understood by observing change of the topology of Fermi surface in the single-particle dispersion (shown in Fig. 3(c)). There have three gaps between branches under the condition $h_{z}>J$. When $\mu$ lies in the lowest gap, the system exhibits a single pair of Fermi points as desired. The increase of $\mu$ tunes the system from $\mathrm{topo-\uppercase\expandafter{\romannumeral1}}$ to $\mathrm{topo-\uppercase\expandafter{\romannumeral2}}$ then back to $\mathrm{topo-\uppercase\expandafter{\romannumeral1}}$ by crossing the critical points between gaps, as shown in Fig .3(b).

\subsection{Observation of Majorana fermions in a harmonic trap.}
To confirm the topological phase transition and observe the appearance of zero-energy MF mode under some appropriate parameter condition, we take into account the trapping potential that is necessary to prevent the atoms from escaping.
Within the LDA, the chemical potential can be thought of as a position dependent function: $\tilde{\mu}(x)=\mu-V(x)$, that continuously decreases away from the center of the trap. Therefore, the critical Zeeman field can be accordingly redefined as
\begin{equation}
h_{c}=\sqrt{\Delta(x)^2+(J\pm \mu(x))^2}.
\end{equation}
Both the local pairing gap and chemical potential decrease away from the trap center. At a small field $h_{z}$, $h_{z}<h_{c1}(x)$ for any position and the whole Fermi cloud is in the conventional superfluid.
Meanwhile, the condition $h_{z}>h_{c2}(x)$ may be always satisfied at an even large Zeeman field.
We can tune the system through an intermediate mixed phase in which topological trivial and non-trivial phase can co-exist.
The mixed phases which appeare naturally in the parabolic trap potential, are illustrated in Fig. 4(a1,b1,c1).
We first consider setting the central chemical potential $\mu(0)$ at a point above all Zeeman gaps that four transition points are all covered by the position-dependent chemical potential (shown in Fig. 4(a1)).
There are five districts along with continuously decrease of the chemical potential from the middle of the trap towards the flanks. If we lower down the chemical potential, less transition points are covered and less topological phases can be distinguished (see Fig. 4(b1,c1)).
All we discussed just depend on the starting point of the chemical potential.

The number of the Majorana zero-energy modes is decided by the amount of spatial interfaces between the different phases discussed above.
The MF is a half of an ordinary fermion and must always come in pairs. Each of the paired states, localized in real space, can be hardly pushed away from $E=0$ by a local perturbation, giving rise to the intrinsic topological stability enjoyed by MFs.
Here we use the numerical method and solve self-consistent BdG equation in real space to explore the MF physics beyond the LDA.
In the presence of a trap $V(x)=m\omega^2x^2/2$,
the BdG equation in real space can be written as $H_{\mathrm{BdG}}(x)\varphi_{\eta}(x)=E_{\eta}\varphi_{\eta}(x)$, where
\begin{equation}
\begin{split}
H_{\mathrm{BdG}}(x)=
\begin{pmatrix}
& H_{S}^{0}(x) & J & 0 & \sigma_{z}\Delta_{1}(x)\sigma_{x}\\
& J & H_{S}^{0}(x) & \sigma_{z}\Delta_{2}(x)\sigma_{x} & 0 \\
& 0 & -\sigma_{z}\Delta_{2}^{\ast}(x)\sigma_{x} & -\sigma_{z}H_{S}^{0\ast}(x)\sigma_{z} & -J \\
& -\sigma_{z}\Delta^{\ast}_{1}(x)\sigma_{x} & 0 & -J & -\sigma_{z}H_{S}^{0\ast}(x)\sigma_{z} \\
\end{pmatrix},
\end{split}
\end{equation}
$\varphi_{\eta}(x)\equiv\left[u_{1\uparrow \eta}(x), u_{1\downarrow \eta}(x), u_{2\uparrow \eta}(x), u_{2\downarrow \eta}(x), v_{2\uparrow \eta}(x),v_{2\downarrow \eta}(x), v_{1\uparrow \eta}(x),v_{1\downarrow \eta}(x)\right]^{T}$
are the Nambu spinor wave functions corresponding to the quasi-particle excitation energy $E_{\eta}$. The single particle Hamiltonian in real-space becomes $H_{S}^{0}(x)=-(1/2m)\partial^{2}/\partial x^{2}+m\omega^{2}x^{2}/2-\mu-h_{z}\sigma_{z}-\alpha i\partial/\partial x \sigma_{y}$. The eigenfunctions satisfy
the normalization condition
\begin{eqnarray}
\int dx \sum_{j=1,2}\left[u^{\ast}_{j\uparrow \eta^{'}}(x)u_{j\uparrow \eta}(x)+u^{\ast}_{j\downarrow \eta^{'}}(x)u_{j\downarrow \eta}(x)+v^{\ast}_{j\uparrow \eta^{'}}(x)v_{j\uparrow \eta}(x)+v^{\ast}_{j\downarrow \eta^{'}}(x)v_{j\downarrow \eta}(x)\right]=\delta_{\eta,\eta^{'}}.
\end{eqnarray}
The order parameter $\Delta_{j}(x)$ and the chemical potential $\mu$ can be determined by the self-consistency equation
\begin{equation}
\Delta_{j}(x)=-g_{1D}/2\sum \left[u_{j\uparrow \eta}v^{\ast}_{j\downarrow \eta}f(E_{\eta})+u_{j\downarrow \eta}v^{\ast}_{j\uparrow \eta}f(-E_{\eta})\right]
\end{equation}
and the number equation $N=\int dx \sum_{j,\sigma}n_{j\sigma} (x)$ with
\begin{equation} n_{j\sigma}(x)=\langle\Psi^{\dagger}_{j\sigma}(x)\Psi_{j\sigma}(x)\rangle=1/2\sum_{\eta}[\left|u_{j\sigma \eta}(x)\right|^2f(E_{\eta})+\left|v_{\sigma \eta}(x)\right|^2f(-E_{\eta})]
\end{equation}
is the local density of $\sigma$ fermions in $j$-th layer, $f(x)=1/(1+e^{x/k_{B}T})$ is the quasi-particle Fermi-Dirac distribution at the temperature $T$.
Similar to the intrinsic symmetry built in the BdG Hamiltonian in momentum space, $H_{\mathrm{BdG}}(x)$ is invariant under the particle-hole transformation: $E_{\eta}\rightarrow-E_{\eta}$ and $u_{j\sigma\eta}(x)\rightarrow v_{j\sigma\eta}^{\ast}(x)$.
To remove this redundancy, a factor of $1/2$ has appeared in the above expressions.
Due to $E_{\eta}=0$, we will immediately have $\Gamma_{0}=\Gamma_{0}^{\dag}$, a zero-energy quasi-particle being its own antiparticle, exactly the defining feature of
a MF.
It is straightforward to show from the BdG Hamiltonian that the wave function of MFs should satisfy either $u_{j,\sigma}(x)=v_{j,\sigma}^{*}(x)$
or $u_{j,\sigma}(x)=-v_{j,\sigma}^{*}(x)$.

We self-consistently solve the gap equation and the number equation for different total number of particles.
$E_{r}/ \omega=50$ is set to assure the trap oscillation frequency is much smaller than the recoil frequency.
Commonly, we use a dimensionless interaction parameter, $\gamma\equiv -m g_{1D}/n_{0}$, to characterize the interaction strength, where $n_{0}$ is the zero-temperature center density of an ideal Fermi gas.
The typical interaction strength is about $\gamma= 3\sim5$ and we take $\gamma=\pi\simeq3.14$ in our calculations.
We have been taken $400$ harmonic oscillators as the expansion functions and find that $N=400$ is large enough to ensure the accuracy of the calculations.

By varying the Zeeman field $h_{z}$ while keeping the chemical potential fixed in the middle of the gap, this way has been proved to be a conventional tool in experiments.
Alternatively, changing the number of particles while keeping the Zeeman field fixed will have the same effect.
Fig. 4(a2,b2,c2) show the mean-field results of energy spectrum with different total particle number.
The emergence of MFs can be clearly revealed by the behavior of the energy spectrum.
At a large total particle number $N_{atom}=200$, eight Majorana zero-energy modes associated with the presence of phase transitions, appear in the middle of the gap (seen in Fig. 4(a2)).
The number of MFs changes as the total number of atoms varies.
We could have six (Fig. 4(b2)) or four (Fig. 4(c2)) and even less MFs corresponding to what we have discussed in Fig. 4(b1,c1).
In Fig. 5, we give a calculation of $h_{c,j}(x)-h_{z}$ for a local uniform cell at position $x$ with the local chemical potential $\mu(x)$ and order parameter $\Delta(x)$.
The local uniform cell would be in the $topo-\uppercase\expandafter{\romannumeral1}$ ($topo-\uppercase\expandafter{\romannumeral2}$) state if $h_{c,1}(x)<h_{z}<h_{c,2}(x)$ ($h_{z}>h_{c,2}(x)$) and the Fermi cloud is in the conventional superfluid with $h_{z}<h_{c,1}(x)$ for any other position $x$.
These areas are highlighted by different colors. At a total particle number $N_{atom}=200$, there are six districts of topological state and therefore we can find four pairs of MFs over the whole system.
By decreasing the total number, the conventional area at the trap center gets shrunk and eventually disappears at about $N_{atom}\sim170$, three pairs of MFs get preserved.
It is obvious that less MFs remain if we future decrease the total number (see Fig. 5(b,c)).
Zero-energy modes just locate around the place with $h_{z}=\sqrt{\Delta(x)^2+(J\pm \mu(x))^2}$, corresponding to the steep slope of the order parameter.
The agreement between BdG and LDA result is reasonable. Both theories predict multi-shell structures.
Their spatial variation clearly identifies that.

\section*{Discussion}
In this work, we investigate the topological properties of bilayer Fermi gases in the presence of SO coupling. 
Two diverse topological phases with $\langle S_{z}(0)\rangle=1,2$ are observed and connected by the TQRT, which have not been discovered previously. 
The topological structure of such system, whether it can be topological trivial or topological non-trivial, depend on not only the combined effect of SO coupling and the Zeeman field, but also the inter-layer tunneling, thus offering more tunability in experiments. 
We give the reliable results by solving the self-consistent BdG equation within a harmonic trap and predict that one can manipulate numerous Majorana zero energy modes associated with interfaces between diverse phases by reforming the phase geometric construction of the system. 
Our results also can be applicable to two dimensional topological superfluids. 
It would be interesting to examine the Chern number that denotes the topology of the matter, and can be used to further test the concept behind our studies. 
The bilayer Fermi gases with SO coupling described here offer an practically realizable platform for further experiments to investigate the topological superfluid and realize and manipulate MFs, providing new insights into quantum computation. 

\section*{Methods}
\subsection{The self-consistent Bogoliubov-de Gennes formalism.}
We work in a complete basis of single-particle wave functions in the harmonic trap $\varphi_{n}(x)$ with energy $\epsilon_{n}=(n+1/2)\omega(n=0,1,2,\ldots)$,
\begin{equation}
\varphi_{n}(x)=\sqrt{1/(a_{h}\pi^{1/2}2^{n}n!)}H_{n}(x/a_{h})\exp(-\frac{x^{2}}{2a_{h}^{2}}),
\end{equation}
here $H_{n}(x)$ is the Hermite polynomial , $a_{h}=\sqrt{1/(m\omega)}$ is the characteristic harmonic oscillator length. Note that spin-orbit coupling working at these basis becomes $-\alpha\partial/\partial x\varphi_{n}(x)=-\alpha/a_{h}(\sqrt{\frac{n}{2}}\varphi_{n-1}(x)-\sqrt{\frac{n+1}{2}}\varphi_{n+1}(x))$.
Then we expand the normalized wave functions as $u_{j,\sigma}(x)=\sum_{n}A_{j,\sigma,n}\varphi_{n}(x)$, $v_{j,\sigma}(x)=\sum_{n}B_{j,\sigma,n}\varphi_{n}(x)$.
according to the orthonormality relation $\int dx \varphi_{n}(x)\varphi_{n'}(x)=\delta_{n,n'}$, we reduce the BdG equation to a matrix diagonalization problem,
\begin{equation}
\begin{pmatrix}
H_{1,\uparrow}^{nn'} & V_{SO}^{nn'} & J\delta_{nn'} & 0 & 0 & 0 & 0 & -\Delta_{1}^{nn'}\\
-V_{SO}^{nn'} & H_{1,\downarrow}^{nn'} & 0 & J\delta_{nn'} & 0 & 0 & \Delta_{1}^{nn'} & 0\\
J\delta_{nn'} & 0 & H_{2,\uparrow}^{nn'} & V_{SO}^{nn'} & 0 & -\Delta_{2}^{nn'} & 0 & 0\\
0 & J\delta_{nn'} & -V_{SO}^{nn'} & H_{2,\downarrow}^{nn'} & \Delta_{2}^{nn'} & 0 & 0 & 0\\
0 & 0 & 0 & \Delta_{2}^{nn'} & -H_{2,\uparrow}^{nn'} & -V_{SO}^{nn'} & -J\delta_{nn'} & 0\\
0 & 0 & -\Delta_{2}^{nn'} & 0 & V_{SO}^{nn'} & -H_{2,\downarrow}^{nn'} & 0 & -J\delta_{nn'}\\
0 & \Delta_{1}^{nn'} & 0 & 0 & -J\delta_{nn'} & 0 & -H_{1,\uparrow}^{nn'} & -V_{SO}^{nn'}\\
-\Delta_{1}^{nn'} & 0 & 0 & 0 & 0 & -J\delta_{nn'} & V_{SO}^{nn'} & -H_{1,\downarrow}^{nn'}\\
\end{pmatrix}
\begin{pmatrix}
& A_{1,\uparrow,n'}\\
& A_{1,\downarrow,n'}\\
& A_{2,\uparrow,n'}\\
& A_{2,\downarrow,n'}\\
& B_{2,\uparrow,n'}\\
& B_{2,\downarrow,n'}\\
& B_{1,\uparrow,n'}\\
& B_{1,\downarrow,n'}\\
\end{pmatrix}
=E\begin{pmatrix}
& A_{1,\uparrow,n}\\
& A_{1,\downarrow,n}\\
& A_{2,\uparrow,n}\\
& A_{2,\downarrow,n}\\
& B_{2,\uparrow,n}\\
& B_{2,\downarrow,n}\\
& B_{1,\uparrow,n}\\
& B_{1,\downarrow,n}\\
\end{pmatrix},
\end{equation}
where the matrix elements are
\begin{equation}
\begin{split}
& H_{j,\uparrow}^{nn'}=(\epsilon_{n}-\mu-h_{z})\delta_{nn'},\\
& H_{j,\downarrow}^{nn'}=(\epsilon_{n}-\mu+h_{z})\delta_{nn'},\\
& V_{SO}^{nn'}=-\alpha(\sqrt{\frac{n'}{2}}\delta_{n,n'-1}-\sqrt{\frac{n'+1}{2}}\delta_{n,n'+1}),\\
& \Delta_{j}^{nn'}=\int dx \varphi_{n}(x)\Delta_{j}(x)\varphi_{n'}(x),\\
\end{split}
\end{equation}
due to the normalization of the quasi-particle wavefunctions, the coefficients of the eigenstate must satisfy the condition $\sum_{n,j,\sigma}(A_{j,\sigma,n}^2+B_{j,\sigma,n}^2)=1$. The same procedure also reduces the order parameter to
\begin{equation}
\Delta_{j}(x)=-g_{1D}/2\sum_{n,n',\eta}A_{j,\uparrow,n,\eta}B^{\ast}_{j,\downarrow,n',\eta}\varphi_{n}(x)\varphi_{n'}(x)f(E_{\eta})
+A_{j,\downarrow,n,\eta}B^{\ast}_{j,\uparrow,n',\eta}\varphi_{n}(x)\varphi_{n'}(x)f(-E_{\eta})
\end{equation}
and the local-density equations
\begin{equation}
n_{j,\sigma}=1/2\sum_{n,\eta}|A_{j,\sigma,n,\eta}|^{2}\varphi^{2}_{n}(x)f(E_{\eta})+|B_{j,\sigma,n,\eta}|^{2}\varphi^{2}_{n}(x)f(E_{\eta}).
\end{equation}
The above equations must be solved together by means of an iterative procedure, which starts from trial functions ($\Delta_{j}(x)$, $\mu$) and converges to the self-consistent solution. All sums in the equations are limited by an energy cutoff $E_{c}=4E_{R}$. This cutoff is required in order to cure the ultraviolet divergences. During the iteration, the order parameters $\Delta_{j}(x)$ are updated, and the chemical potential $\mu$ is also adjusted slightly to enforce the number-conversation condition. The self-consistent iterative procedure is over until final convergence is reached.

\begin{addendum}

\item [Acknowledgement]

We thank An-Chun Ji and Qing Sun for their early contribution to this project and are grateful to KashifAmmar Yasir and Xiao-Xi Lv for valuable discussions. This work was supported by the NSFC under grants Nos. 11434015, 61227902, 61378017, NKBRSFC under grants Nos. 2012CB821305, SKLQOQOD under grants No. KF201403, SPRPCAS under grants No. XDB01020300.

\item [Author Contributions]
All authors planned and designed theoretical and numerical studies.
All contributed in completing the paper.

\item [Competing Interests]
The authors declare that they have no competing financial interests.
\item [Correspondence]
Correspondence should be addressed to Wu-Ming Liu (email: wliu@iphy.ac.cn).
\end{addendum}
\newpage
\bigskip
\textbf{Figure 1 The experimental setup for spin-orbit coupled bilayer Fermi gases.} (\textbf{a}) Schematics diagram for the bilayer system of $\leftidx{^{40}}{\mathrm{K}}$ Fermi gases. A double well optical lattice that has two local minima in a unit cell, is formed by the interference of two pairs of counter-propagating laser beams. 
In this case, there is a resulting double-well periodic potential along $\hat{z}$-axis, which can allows the inter-well tunnelling $J$. 
Within a unit cell, the barrier height and relative depth of the double well can be regulated by changing the intensity or relative phase of laser standing waves. 
Two Raman lasers counter-propagate along $\hat{x}$-axis are used to generate SO coupling with the recoil momentum $k_{r}=k_{0}\sin(\theta/2)$, where $k_{0}=2\pi/\lambda$ and $\theta=\pi$ is the angle between two Raman lasers. 
(\textbf{b}) Effective coupling between four atomic levels $|j,\sigma\rangle$ ($j=1,2, \sigma=\uparrow, \downarrow$) in a double-well trap, where $J$ denotes the inter-layer tunneling without spin flipping, and $\Omega$ represents the intra-layer spin-orbit coupling induced by the Raman lasers. 
(\textbf{c},\textbf{d}) show the evolution of single-particle energy spectra as a function of the Zeeman field $h_{z}$. 
Both without (\textbf{c}1)-(\textbf{c}3) and with (\textbf{d}1)-(\textbf{d}3) in-plane Zeeman field are considered. 
Without in-plane Zeeman field, the band structure is symmetric around $k=0$, which supports the superfluid pairing between atoms with opposite momentum. 
The band structure becomes asymmetric and a finite total momentum pairing may occurs when $h_{y}\neq0$. 
The colored energy curves correspond to different Zeeman field $h_{z}$. 
It is worth noting that the crossover between $E_{1,-}$ and $E_{2,+}$ disappears and open a gap between the two. when $h_{z}>J$. 
The tunneling is given by $J=0.6E_{r}$, where $E_{r}$ is the recoil energy, chosen as the natural energy unit. 

\newpage
\bigskip
\textbf{Figure 2 Mean-field topological phase diagram at a given chemical potential.}
(\textbf{a}) Topological phase diagram on the $\Delta$-$h_{z}$ plane with  ($\mu/E_{r}=0.5$, $J/E_{r}=0.3$): we can distinguish three different topological regions from the behavior of the energy gap $E_{g}$. 
The white dotted lines represent phase boundaries where the energy gap equal zero, and the colors ranging from dark to red describe the values of $E_{g}$.
(\textbf{b}) Plot of the energy gap $E_{g}$ and the spin polarization $\langle S_{z}(k)\rangle $ at zero momentum  with respect to the Zeeman field $h_{z}$. 
Here, the order parameter is chosen as $\Delta/E_{r}=0.5$, the tunneling between two layers as $J/E_{r}=0.5$ and the chemical potential as $\mu/E_{r}=0.5$. 
There is a jump in the spin polarization at $k=0$, occurring precisely at the phase interface, where the topology of the Fermi gases changes. 
The phases are labelled with different colors: topological trivial superfluid (\textbf{trivial}, blue) with $h_{z}<h_{c1}$, a new type of topological superfluid ($\mathbf{topo-\uppercase\expandafter{\romannumeral1}}$, gray) with $h_{c1}<h_{z}<h_{c2}$, and normal topological superfluid ($\mathbf{topo-\uppercase\expandafter{\romannumeral2}}$, pink) with $h_{z}>h_{c2}$. The red triangles mark the critical values of Zeeman field $h_{c}=\sqrt{\Delta^2+(J\pm \mu)^2}$.
(\textbf{c})-(\textbf{f}) show the quasi-particle spectrum in momentum space at different Zeeman field $h_{z}$. As the Zeeman field increases, the gap closes and opens when $h_{z}\sim0.5E_{r}$, and then re-closes at the point of $h_{z}\sim1.1E_{r}$. The system evolves from a conventional superfluid to a $\mathbf{topo-\uppercase\expandafter{\romannumeral1}}$ superfluid and finally to a $\mathbf{topo-\uppercase\expandafter{\romannumeral2}}$ state. Other parameters are the same as in (\textbf{b}).

\newpage
\bigskip
\textbf{Figure 3 Phase diagram and the evolution of spin-orbit coupled bilayer Fermi gases with the increasing chemical potential.}
(\textbf{a}) Topological phase diagram along with the energy gap, is shown on $\Delta$-$\mu$ plane.
With increasing the strength $\Delta$ of the pairing potential, the area of $\mathbf{topo-\uppercase\expandafter{\romannumeral2}}$ gets smaller and finally disappear above a threshold, $\Delta=\sqrt{h_{z}^2-J^2}$. If we move on and reach the limit $h_{z}^2-\Delta^2=0$, the full topological trivial occurs in the absence of topological transition.
(\textbf{b}) The energy gap $E_{g}$ is plotted as a function of the chemical potential $\mu$ with $\Delta/E_{r}=0.5$. The colored regions are determined by gap closing equation: $h_{c}=\sqrt{\Delta^2+(J\pm \mu)^2}$, as we have used in Fig .2. There are five phase areas indicated by different colors and separated by the critical values of the chemical potential $\mu_{c,i}$ with $(i=1,2,3,4)$, marked by the red triangles. The Zeeman field is fixed at $h_{z}/E_{r}=1$, and the tunneling strength is taken as $J/E_{r}=0.4$.
(\textbf{c}) An illustration of the topological phase transition at different chemical potentials for small $\Delta$. The single particle energy dispersion is drawn by solid lines and the dashed line is the chemical potential. It is found that the system enter topological phase when the chemical potential is tuned to lie inside a Zeeman gap.

\newpage
\bigskip
\textbf{Figure 4 bilayer Fermi gases with synthetic spin-orbit coupling in a harmonic trap.}
The left panel gives a rough configuration of bilayer Fermi gases in a harmonic trap for different chemical potential within the local-density approximation (LDA). MFs just appear around the interface between different regions, approximately where the chemical potential satisfy the critical gap closing equation: $h_{c}=\sqrt{\Delta^2+(J\pm \mu)^2}$. When the chemical potential at the center of the trap $\mu$ is set to be so large that all four branches can be filled, as shown in (\textbf{a}1). We will find four pairs of $\mu_{c,i}$ from the center to the edge of the trap. (\textbf{b}1) and (\textbf{c}1) give the results with lower chemical potential, less pairs of $\mu_{c,i}$ can be detected in the trap. The right panel shows the quasi-particle excitation spectrum $E_{\eta}$ (in units of $E_{R}$), calculated within the self-consistent BdG approach, at $N_{atom}=200$, $N_{atom}=150$ and $N_{atom}=120$, respectively. The number of Majorana zero modes changes as the total number of atoms varies. We therefore could have eight (\textbf{a}2), six (\textbf{b}2) and four (\textbf{c}2) Majorana fermions, corresponding to what we have discussed in the left panel. Here $\gamma=\pi$, $h_{z}/E_{r}=0.7$ and $J/E_{r}=0.2$.

\newpage
\bigskip
\textbf{Figure 5 Spatial dependence of the critical Zeeman fields $h_{c,1}(x)-h_{z}$, $h_{c,2}(x)-h_{z}$ and the superfluid order parameter $\Delta(x)$ at different phases.}
In accord with Fig. 4, we present $h_{c,1}(x)-h_{z}$, $h_{c,2}(x)-h_{z}$ and $\Delta(x)$ for a Fermi gas of (\textbf{a}) $N_{atom}=200$ , (\textbf{b}) $N_{atom}=150$, and (\textbf{c}) $N_{atom}=120$ fermions in trap at zero temperature. The symmetry around $x=0$ originates from the harmonic trapping geometry. The phases distinguished by the local topological phase condition, are highlighted with different colours. The dots in each subgraphs label Majorana fermions associated with interfaces between different phases, where the Zeeman filed cross the lines. By decreasing $N_{atom}$, the number of Majorana zero modes changes with the deformation of topological phase structure. From (\textbf{a}) to (\textbf{b}), the \textbf{trivial} area in the bottom of the trap shrinks and vanishes with disappearance of two Majorana fermions. Therefore, $\mathbf{topo-\uppercase\expandafter{\romannumeral1}}$ occurs at the centre of the well. From (\textbf{b}) to (\textbf{c}), similar to the above case, $\mathbf{topo-\uppercase\expandafter{\romannumeral1}}$ in the middle of the trap disappears and the number of Majorana fermions becomes 2 pairs, corresponding to what we have shown in Fig. 4.

\end{document}